\documentclass[reprint,superscriptaddress,showpacs,preprintnumbers,amsmath,amssymb,showkeys]{revtex4-1}
\usepackage{graphicx}
\usepackage{amsmath}

\usepackage[final]{hyperref}
\usepackage{amssymb}
\usepackage{textcomp}

\begin{document}

\preprint{published in: Nanotechnology 20, 455501 (2009). \href{http://dx.doi.org/10.1088/0957-4484/20/45/455501}{http://dx.doi.org/10.1088/0957-4484/20/45/455501}}
\title{Superconducting NbN detector for neutral nanoparticles}

\author{Markus Marksteiner}
\affiliation{Faculty of Physics, University of Vienna, Boltzmanngasse 5,
A-1090 Vienna, Austria} 
\author{Alexander Divochiy}
\affiliation{Department of Physics, Moscow State Pedagogical
University, M. Pirogovskaya Street 1, Moscow 119992, Russia}
\author{Michele Sclafani}
\author{ Philipp Haslinger}
\affiliation{Faculty of Physics, University of Vienna, Boltzmanngasse 5,
A-1090 Vienna, Austria}
\author{Hendrik Ulbricht}
\affiliation{School of Physics and Astronomy, University of Southampton, Highfield, Southampton, SO171BJ, United Kingdom}
\author{Alexander Korneev}
\author{Alexander Semenov}
\author{Gregory Gol'tsman}
\affiliation{Department of Physics, Moscow State Pedagogical
University, M. Pirogovskaya Street 1, Moscow 119992, Russia}
\author{Markus Arndt}
\email{markus.arndt@univie.ac.at}
\homepage{http://www.quantumnano.at}
\affiliation{Faculty of Physics, University of Vienna, Boltzmanngasse 5,
A-1090 Vienna, Austria}

\date{\today}
\begin{abstract}
We present a proof-of-principle study of superconducting single
photon detectors (SSPD)  for the detection of individual neutral
molecules/nanoparticles at low energies. The new detector is
applied to characterize a laser desorption source for
biomolecules and it allows to retrieve the arrival time
distribution of a pulsed molecular beam containing the amino acid
tryptophan, the polypeptide gramicidin as
well as insulin, myoglobin and hemoglobin. We discuss the experimental
evidence that the detector is actually sensitive to isolated
neutral particles.
\end{abstract}

\maketitle
\section{Introduction}
The detection of isolated neutral molecules and nanoparticles in the
gas phase is both a necessity and a challenge for many experiments
that range from physical chemistry over environmental
monitoring~\cite{Noble2000a} to the foundations of physics. Our own
work was originally motivated by matter wave interferometry with
massive molecules\,\cite{Schollkopf1996a, Arndt1999a,
Hackermuller2003a} and applications in molecule
metrology\,\cite{Gerlich2008a, Ulbricht2008a}. The extension of such
experiments to higher masses requires also improved methods for
detecting neutral nanoparticles. While ionization techniques are
routinely used for particles up to about 2000\,Da, postionization of
organic molecules beyond that mass has remained a significant
challenge\,\cite{Schlag1992a,Becker1995a}. Recent experiments
observed photoionization of tryptophan-metal complexes and
nucleotide clusters up to 6000\,Da
\,\cite{Marksteiner2008a,Marksteiner2009a}. But for the majority of
high-mass biomolecules this method seems to be precluded.

Hyperthermal surface ionization~\cite{Amirav1991a} was shown
to allow the detection of some neutral molecules, with insulin currently
setting the mass record\,\cite{Weickhardt2003a}.

Modern nanofabrication technologies also allow to build nanoscale
oscillators which change their resonance frequency when their mass is augmented by even a single molecule. Such
cantilever based detectors\,\cite{Ilic2004a,Li2007a} have rather a lower than an upper mass limit. They currently reach a sensitivity of below 200\,Da~\cite{Jensen2008a}. First proof-of principle mass spectrometer applications achieved the capability to detect single proteins~\cite{Naik2009a}.

Whenever mass cannot be measured directly, bolometer
detectors\,\cite{Low1961a,Kraus1996a, Enss2008a} may
convert molecular energy first into sensor temperature and then into an electrical signal. However, the translational
energy of a single amino acid, such as tryptophan, does not exceed 0.3\,eV, even at a molecular velocity of 500\,m/s.
This is why the first bolometers~\cite{Cavallini1971a} still operated with a minimum detection threshold of about $10^7$\,molecules per second.
Superconductors were suggested as promising sensors\,\cite{Andrews1946a,Kraus1996a,Booth1996a} since their conductivity changes strongly with temperature in the vicinity of the phase transition edge.

The implementation of superconducting tunneling junctions (STJ) made it
possible to detect charged individual molecules\,\cite{Frank1996a,Twerenbold1996b, Esposito2002a}. This is
interesting for mass spectrometry because the STJ response depends on the particle's energy\,\cite{Twerenbold2001a}. This allows to combine the mass discrimination of a time-of-flight spectrometer with a detector whose efficiency remains constant over a wide mass range.
Tunneling junction detectors require, however, cooling
well below 4\,K and up to now they were only used for recording either ensembles of slow neutral particles or for detecting individual but energetic charged particles.

In this letter we present our first experimental evidence that a
combination of both is feasible, i.e. a detector for single
neutral molecules of low kinetic and low internal energy. Our
nanowire detector was originally fabricated as a
superconducting single photon detector. Its
sensitivity to single photons was demonstrated across the entire
spectrum from UV to mid-IR, with quantum efficiencies up to 30\,\%
\,\cite{Goltsman2001a,Semenov2001a,Verevkin2002a,Korneev2005a,
IEEE07}. Before we started the experiments, it was far from
obvious that such a device would also be sensitive to slow
nanoparticles. In the following we discuss the acquired evidence
that this detector is capable of recording the incidence of
isolated neutral biomolecules.

\section{Experimental setup}
The entire experiment consists of a pulsed molecular beam,
a free flight trajectory in high vacuum and the superconducting nanowire detector in a differentially pumped helium cryostat, about 76\,cm behind the source.

\subsection{Superconducting detectors}
Two different types of detectors were tested: Superconducting single photon counting devices (SSPDs) and superconducting bolometers (SBs).

The SSPDs were fabricated by depositing a NbN film of 3.5..4\,nm thickness on a sapphire substrate. We tested chips with an open area
of either $10\times10\,\mathrm{\mu m}^2$ or
$20\times 20\,\mathrm{\mu m}^2$. The 100..120\,nm wide
superconducting wire meanders on the surface with a filling
factor of 60\,\%.
The critical temperature of NbN is
T$_c$=10..11\,K and the critical current density
amounts to $j_c$=3..5 $\times$\,10$^6$\,A/cm$^2$.

\begin{figure}[tbp]
   \centering
   \includegraphics[width=8.5cm]{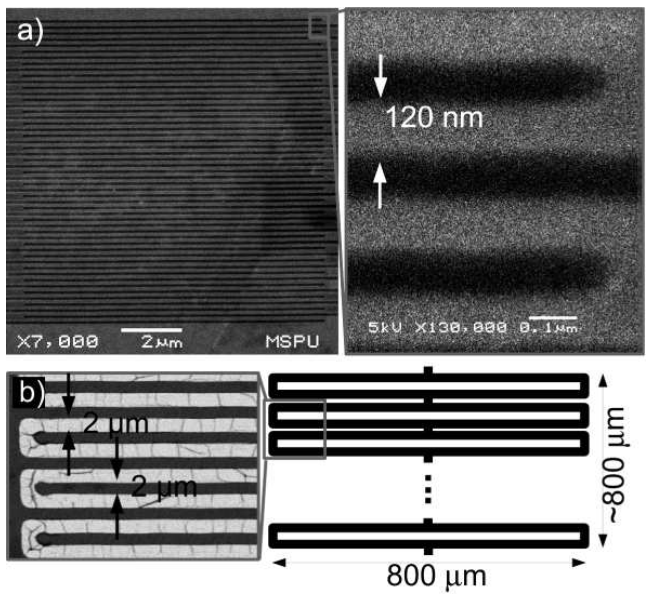}
   \caption{(a) Electron microscopy image of the
   $10\mathrm{\mu m}\times10\mathrm{\mu m}$ sensitive SSPD element.
   The NbN film appears is colored in grey. (b) Sketch (right) and SEM image (left) of the superconducting NbN bolometer. The bolometer strips and gaps are about 2~$\mu$m wide.}
    \label{abb:figure1}
  \end{figure}

\begin{figure*} [htb]
   \centering
   \includegraphics[width=\columnwidth]{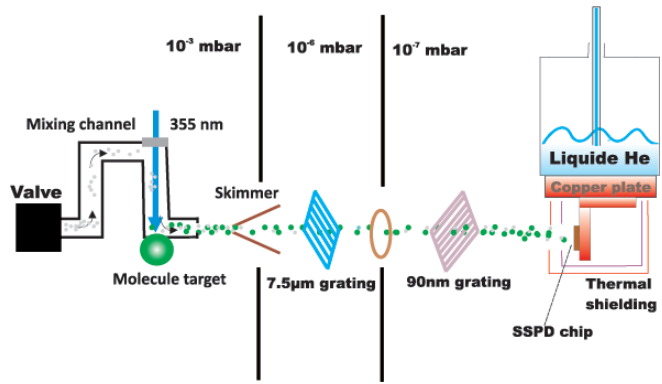}
   \caption{A pulsed laser desorbs biomolecules, which are then entrained by a
   supersonically expanding noble gas jet. After two-fold mechanical filtering the molecular beam hits the superconducting NbN detector. The impact of molecules releases a voltage pulse which is stored by an oscilloscope and a timer card.  }
   \label{abb:figure2}
  \end{figure*}

The SSPD fabrication process was described in detail
in\,\cite{FabricationOf}. In brief, NbN superconducting films were
deposited on R-cut sapphire substrates by DC reactive magnetron
sputtering in an Ar and N$_2$ mixture. The film was patterned by
direct electron beam lithography and reactive ion etching. Gold
contacts were added using photolithography and wet etching.
In Figure~\ref{abb:figure1} we show electron microscopy images of the sensitive element.
We operate the device in a liquid helium bath cryostat at 4.2\,K and
apply a DC bias current slightly smaller than the critical current.
The signal is capacitively coupled from the chip to the oscilloscope (see Figure\,\ref{abb:figure2}).

The detection mechanism may be understood as
follows: When a molecule hits the film surface, it creates
high-energy acoustic phonons. These phonons are rapidly absorbed by
the electron subsystem of the film due to their short inelastic
mean free path. Excess quasiparticles are then created which, in turn,
dispose of their energy by the emission of second generation
phonons. This triggers an avalanche multiplication cascade. The
process is similar to what happens during photon detection.
The distinctive difference lies in the first step: the photon detection
cascade starts from a single high-energy quasiparticle created by
the photon. The dynamics of the subsequent stages is determined only
by the absorbed energy.

When the energy of the quasiparticles
decreases to a value around 10~K, the electron-electron interaction
becomes more efficient for the multiplication of the quasiparticles
than the electron-phonon interaction. Due to this fact, the main
part of the energy, that is initially deposited in the film remains in the quasiparticle subsystem. At the end of
the cascade, a hot spot of excess quasiparticles is formed and
the supercurrent is forced to flow around the new
normal-conducting area. If the hot spot
is sufficiently large the current density in the `sidewalks' increases
beyond the critical current density. This results in a short breakdown of
superconductivity across the entire width of the nanostripe and in a voltage
pulse that can be easily detected\,\cite{Goltsman2001a, Semenov2001a, Verevkin2002a,
Korneev2005a}. For photons, a typical pulse response lasts over
10\,ns.

The second detector type that we tested was a classical
superconducting bolometer. These chips were made from the same
NbN film, again using photolithography and wet etching.
Figure~\ref{abb:figure1}(b) presents the sketch of its
sensitive element and an SEM image.
The bolometer chip is working at the critical temperature $T_c$.
The additional energy that is delivered to the surface by the impacting molecules may heat the superconductor above $T_c$ and cause a voltage peak. The incident energy has to sufficiently high to induce
the required temperature change. Because of the large width of the
stripes this condition can often not be met by a single molecule alone
and the sensors responds only to many simultaneously impacting particles.
With this second detector type we were not able to detect any molecular
signal in our experiments. This type of superconducting wide-area bolometer is therefore not further discussed in the following where we rather focus on our explorations of nanostructured SSPD chips.
\subsection{Molecular beam source}  The details of our
laser desorption source have already been described
elsewhere\,\cite{Marksteiner2008a, Marksteiner2009a}:
Organic molecules were laser desorbed by a Nd:YAG laser beam
(355\,nm, 5\,ns, 6..10\,mJ), which was focused to a spot
on a pressed powder sample
(Figure\,\ref{abb:figure2}).
The desorbed molecules are cooled and entrained by
a jet of helium gas that fills the mixing channel
before it exits through a 1\,mm opening into a vacuum of 10$^{-3}$ \,mbar. About 2\,cm behind the mixing channel the beam passes a skimmer of 1\,mm diameter, which separates the source from a differential pumping stage. In
this second chamber the molecular beam is filtered by a copper mesh
with 7.5\,$\mathrm{\mu m}$ openings. This microstructure is used to reject grains of powder that might be ejected during the ablation process.

The stream of single molecules as well as possibly a background of microscopic particles leaves the second pumping stage through a 1\,cm diameter opening
into the detection chamber where the SSPD chip is
attached to a Helium bath cryostat. The overall distance from
the desorption spot to the superconducting chip is 76\,cm.
For some experiments we added a second mechanical filter:
a SiN$_{x}$ line grating with a period of 266\,nm and
openings as small as 90\,nm was attached to the entrance window of the
cryostat.
This addition further limited the transmitted particle
size and also helped extending the life time of the SSPD chips, which
was otherwise strongly affected by the accumulation of molecular
material. Even with the additional filter in place, the active time
of an individual chip was limited to about 20.000 desorption shots,
time after which a layer of molecules had covered the surface and made it
insensitive. It is known~\cite{Reiger2007a} that  SSPD chips can even be used to resolve the energy of incident photons. Similar energy dependent measurements with molecules were, however,  still impeded in our present proof-of-principle study by the time-varying surface coverage.

\section{Results}
In order to explore its detection capabilities the SSPD chip was
placed into the laser desorbed biomolecular beam. The first
experiments were performed with a mixture of several molecules. It contained 0.2\,g myoglobin (17\,kDa),
0.3\,g $\mathrm{\beta}$-carotene (537\,Da), 0.3\,g insulin
(5.8\,kDa), 0.25\,g  bovine serum albumin (BSA,66\,kDa), and 0.5\,g cellulose of unspecified chain length to mechanically bind the other
components.

\begin{figure}[tbp]
   \centering
   \includegraphics[width=0.6\columnwidth]{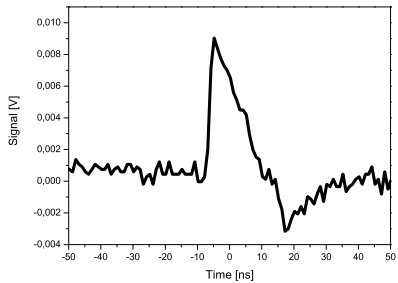}
   \caption{A typical individual peak, that is attributed to the impact of
   neutral molecules on the chip. The
  20$\times$20$\mathrm{\mu}m$ SSPD chip was biased with a current of 19.5\,$\mathrm{\mu}A$. The signal was amplified by 20\,dB.}
   \label{abb:figure3}
  \end{figure}

In this first test we used a $20 \times 20\,\mathrm{\mu m^2}$ SSPD
chip and in Figure\,\ref{abb:figure3} we show a typical individual detection
event from the desorbed molecule mixture. The peak is about 10\,ns
wide (FWHM) and indicates a high temporal detector resolution also
for neutral nanoparticles.

We performed several tests to corroborate the evidence for the
SSPD's sensitivity to isolated molecules and to exclude possible other reasons
for the observed signals such as for instance the co-propagating seed gas or co-desorbed cellulose:

The influence of the rapidly expanding seed gas can be tested by
switching off the desorbing laser beam. We searched for signs of
the expanding helium carrier gas pulse alone and the complete
absence of any signal in this setting indicates that the SSPD
chip is not capable of detecting individual helium atoms. The
same is also true for all of the heavier noble gases such as
neon, argon, krypton or xenon. None of them showed any detectable
signal under our experimental conditions.

One might speculate that the higher kinetic energy of the more
massive biomolecules could be outweighed by the larger number of
lighter noble gas atoms. This argument would be supported by the
fact that seed gas atoms must be much more abundant than the laser implanted biomolecules --
otherwise supersonic expansion would never occur. However, since
no signal was detected for the pure noble gas beams alone, a
collective effect by the dense gas jet can be excluded. Since the
biomolecular beam is more dilute than the carrier gas, a
collective effect of organic particles is even less likely.

This finding is in variance to that for classic bolometers where the
incidence of many particles was actually required to trigger a
signal~\cite{Cavallini1971a}. It has, however, to be noted that these
detectors were not nanostructured and they were exploiting a
different mechanism. Our result gives thus first
evidence that the SSPD chip is indeed not sensitive to the intense
particle flux of atoms but rather to the local energy density of
single complex nanoparticles.

As mentioned before, the molecules were always admixed with
cellulose to achieve mechanical stabilization of the sample. In
order to separate matrix signals from analyte signals we also
performed a separate desorption experiment with pure cellulose
powder alone. The complete absence of any measurable signal
indicates again that the ablated matrix particles do not contribute
any background in the SSPD counter.
In order to enable a more quantitative evaluation we switched to a pulse counting mode and recorded time-of-flight curves for various
experimental settings. The molecular velocity may differ from that of a free supersonic expansion, since the
molecules can be delayed inside the gas mixing channel before they
exit. This delay may lead to an underestimation of the actual
velocity. The velocities and kinetic energies below are therefore reasonable lower limits.

\begin{figure}[htb]
   \centering
   \includegraphics[scale=0.9]{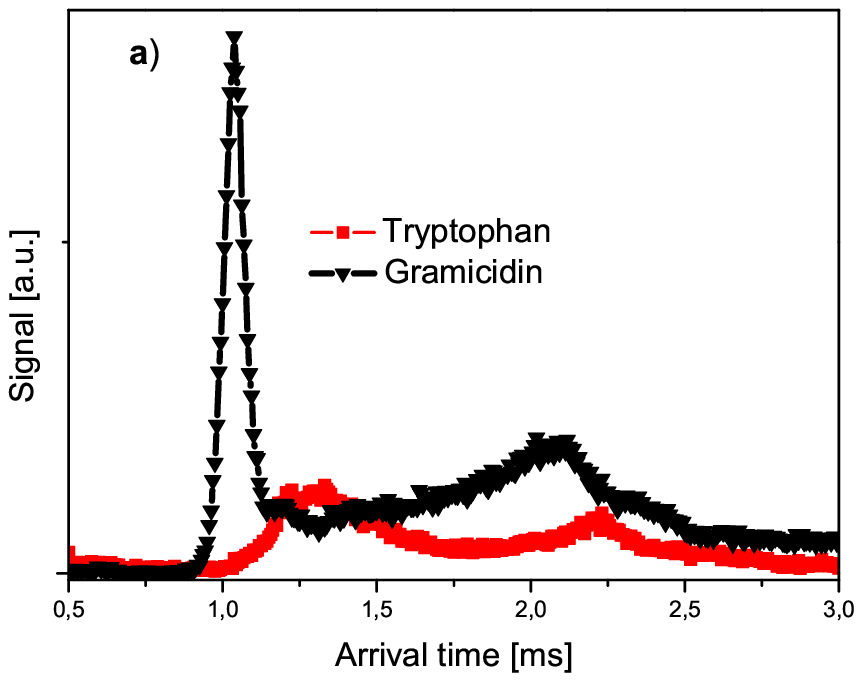}
   \includegraphics[scale=0.9]{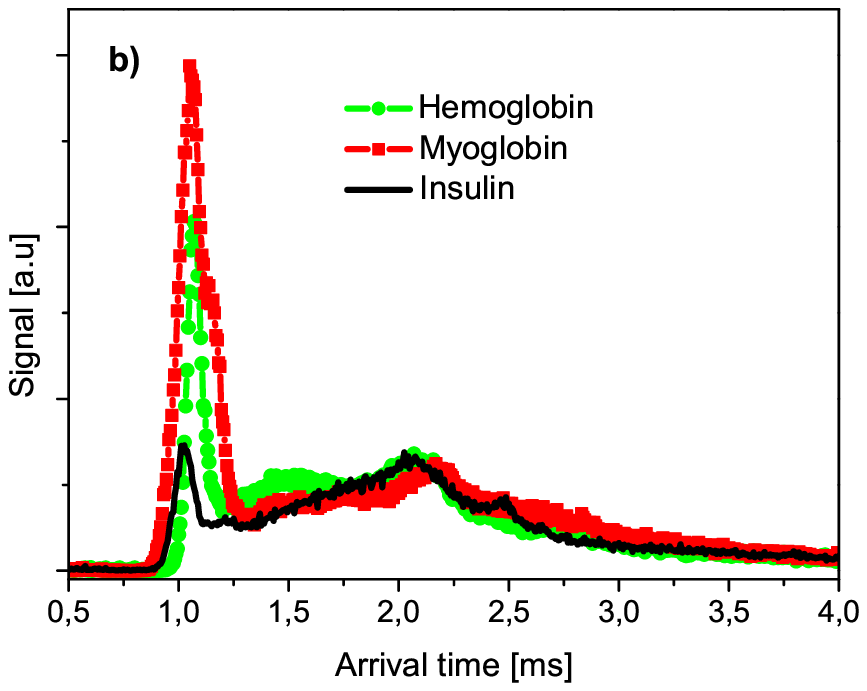}
   \caption{A) Arrival time distribution of tryptophan and
   gramicidin. B)
   arrival time distribution of myoglobin, insulin and hemoglobin. All molecules were detected using the 20$\times$20\,$\mathrm{\mu}m$ SSPD chip.}
   \label{abb:figure4}
  \end{figure}

In all the following experiments the samples contained only
biomolecules from one species mixed 1:1 with cellulose. In
Figure\,\ref{abb:figure4}a we show the arrival time distribution for
a tryptophan (204\,Da) and gramicidin (1.9\,kDa) sample,
respectively. Figure\,\ref{abb:figure4}b depicts the distributions
for insulin, myoglobin and porcine hemoglobin (66\,kDa) sample.
All curves in Figure\,\ref{abb:figure4} were recorded using the same
$20 \times 20\, \mathrm{\mu}m$ SSPD chip, the same discriminator
level, a bias current of 20$\pm$1\,$\mathrm{\mu}A$ and two
particle filters in the beam line, i.e. a 7.5\,$\mathrm{\mu m}$ mesh
as well as the 90\,nm SiN filter. The opening time of the valve was set to 700\,$\mu$s for all recorded curves in Figure\,\ref{abb:figure4}, except for the one of tryptophan as discussed below.
The arrival time distributions in Figure\,\ref{abb:figure4} show a
double structure which is a result of the particular valve setting in these experiments.

The agreement of the detected arrival times with the expected
flight times may already be interpreted as a good indication for
the detection of neutral particles. It is, however, desirable to
corroborate this statement by complementary measurements which
must rely on alternative detectors. As they are not readily
available in the mass range beyond 2000\,Da, where ionization
detectors start to fail\,\cite{Schlag1992a, Becker1995a} and
nanomechanical detectors are not yet commercially available, it
is also still open whether high mass molecules (e.g. hemoglobin)
can survive the desorption process as intact particles or whether
the observed signals are rather caused by smaller neutral
fragments generated in the source.

\begin{figure}[htbp]
   \centering
   \includegraphics[scale=0.8]{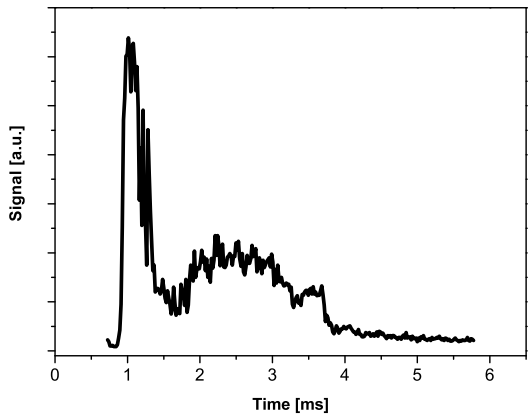}
   \caption{Arrival time distribution of tryptophan recorded via laser
   postionization and time-of-flight (TOF) mass spectrometry. The shape of the TOF-distribution reproduces the distribution recorded by the superconducting chip (see figure\,\ref{abb:figure4}). }
   \label{abb:figure5}
  \end{figure}
This is why our first checks were focused on characterizing the
molecular beam source for tryptophan and gramicidin, where it is
known\,\cite{Koster1992a, Arps1989a} that VUV laser ionization is
soft and capable of detecting isolated molecules as well as larger
clusters\,\cite{Marksteiner2008a, Marksteiner2009a}. A F$_2$ laser (157\,nm, 5\,ns, up to 3\,mJ) is here combined with time-of-flight mass spectrometry
(TOF-MS) to reveal the arrival time and mass distribution of all
molecules emerging from the source.

Figure\,\ref{abb:figure5} shows a representative arrival time
distribution for tryptophan, measured using TOF-MS and the same
source settings as in the SSPD experiments. Only the flight distance
between source and detector was shortened to 0.5\,m. This reduces
also the molecular flight times.

\begin{figure}[tbp]
   \centering
   \includegraphics[scale=0.8]{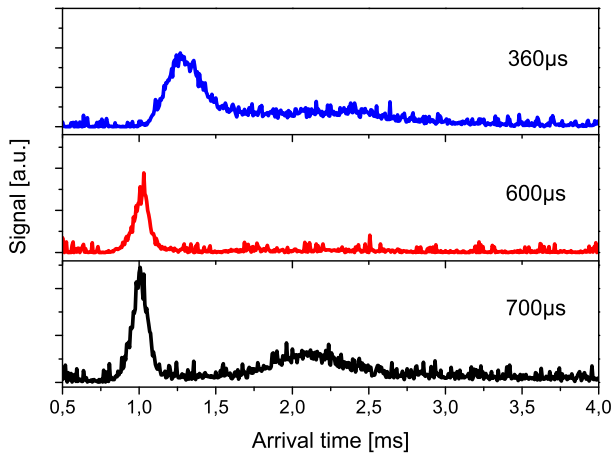}
   \caption{Arrival time distribution of gramicidin for different
   valve opening times, using a  $10\times 10\,\mathrm{\mu m}$
   SSPD chip. A reduction of the valve opening time leads to shifted arrival time distributions (slower molecules).}
    \label{abb:figure6}
  \end{figure}

The arrival time distribution, recorded in photoionization TOF-MS at the mass of the single monomer (Figure\,\ref{abb:figure4}) has the same structure as the signals recorded by the SSPD (Figure\,\ref{abb:figure5}). Small differences in the arrival times can be assigned to an accidental reduction of the valve
opening time which influences both the pressure inside the mixing channel and the velocity of the expanding molecules.
This is supported by Figure\,\ref{abb:figure6}
which depicts the arrival time distribution for gramicidin for three
different valve opening times recorded by the SSPD chip. For short
pulse times (Figure\,\ref{abb:figure6}, top) the arrival time
distribution resembles the one recorded for tryptophan in
\ref{abb:figure4}a, where the molecule arrival time was delayed. Interestingly, valve times around 600\,$\mu$s do
not show the double structure, as can be seen in
Figure\,\ref{abb:figure6}, center. We chose a slightly higher
opening time of 700\,$\mu$s (Figure\,\ref{abb:figure6}, bottom) for
our experiments. This causes a double peak in the arrival time but
also adds to the signal.

From these tests we gather the general insight that the arrival time distributions are rather identical for both the SSPD and the ionization detector. They are both consistent with the arrival of individual molecules.
The absence of any signal related to the individual carrier gas
atoms is a hint to an energy threshold in the SSPD which can only
be overcome by sufficiently massive and energetic molecules. In our
experiments the least energetic molecules detected
by the SSPD were tryptophan particles in the velocity band of 300..500\,m/s, i.e. with a kinetic energy of 100..300\,meV if we assume to see isolated molecules.

Compared to that value, all rare gas beams have still too little kinetic
energy to be detected. Helium at 800\,m/s reaches only 10\,meV and xenon at 250\,m/s would only attain 40\,meV -- still well below the value for tryptophan, which carries also internal energy in addition.

\section{Conclusion}
The present experiments are just a promising start of an interesting
journey into single neutral molecule detection using superconducting
single photon detectors.
Our experiments give good first evidence that
SSPDs can be used to register the incidence of neutral single nanoparticles.

As of today there is no efficient easy-to-implement way of
detecting neutral large proteins to cross-check our results with
individual, isolated insulin, hemoglobin or myoglobin. But we see
a good consistence between the SSPD results and photoionization mass spectrometry in the flight times for tryptophan and gramicidin.

One might furthermore ask whether a high \emph{internal} excitation
may also add sufficient energy to the chip, which will be tested in future experiments by systematically varying the internal temperature of the molecules.

The possibility of detecting more massive neutral and labile
molecules is promising for many applications in physical chemistry
and also for matter wave interferometry. Even if the SSPD method,
cannot (yet) discriminate between different masses, de Broglie
interferometry itself has been shown to be capable of discriminating
different molecular properties\,\cite{Gerlich2008a} and experiments
with clean and mass selected sources would only require a good
sensitivity to the existence, not to the mass of the particle.

The technology certainly requires further development and
exploration but it may allow us to close a 'detector loophole' for
particles which are too complex to be efficiently photoionized
and yet too small to be well detected by other means.

\section*{Acknowledgments}
This work was supported by the Austrian Science Funds FWF within the
Wittgenstein program Z149 as well as by the Italian Fondazione
Angelo Della Riccia. We thank Sanofi-Aventis Inc. for the donation
of pure insulin powder. This work was also supported by the Russian
Agency of Science and Innovation (contract 02.513.11.3446) and the
Russian Foundation for Basic Research (grant 09-02-12364).

\section*{References}

\bibliography{references}

\end{document}